\documentclass[prl,twocolumn,calc,epsfig]{revtex4-1}

\usepackage{graphics}
\usepackage{amsfonts}
\usepackage{amsmath}
\usepackage{epsfig}

\begin{document}

\title{Motion of a Solitonic Vortex in the BEC-BCS Crossover}
\author{Mark J.H. Ku, Wenjie Ji, Biswaroop Mukherjee, Elmer Guardado-Sanchez, Lawrence W. Cheuk, Tarik Yefsah, and Martin W. Zwierlein}

\affiliation{MIT-Harvard Center for Ultracold Atoms, Research Laboratory of Electronics, and
Department of Physics, Massachusetts Institute of Technology, Cambridge, Massachusetts 02139, USA}

\begin{abstract}
We observe a long-lived solitary wave in a superfluid Fermi gas of $^6$Li atoms after phase-imprinting.
Tomographic imaging reveals the excitation to be a solitonic vortex, oriented transverse to the long axis of the cigar-shaped atom cloud. The precessional motion of the vortex is directly observed, and its period is measured as a function of the chemical potential in the BEC-BCS crossover. The long period and the correspondingly large ratio of the inertial to the bare mass of the vortex are in good agreement
with estimates based on superfluid hydrodynamics that we derive here using the known equation of
state in the BEC-BCS crossover.
\end{abstract}

\maketitle

Solitary waves that do not spread as they propagate are ubiquitous in non-linear systems, from classical
fluids and fiber optics to superfluids and superconductors. Their motion plays a central role in phenomena as
diverse as the conductivity of polymers~\cite{Heeger1988SSH}, the finite resistance of superconductors
in a magnetic field~\cite{Halperin2010Resistance}, pulsar
glitches~\cite{Anderson1975Pulsar} and likely the formation of the
Universe~\cite{kibb80}.

\begin{figure}
    \begin{center}
    \includegraphics[width=89mm]{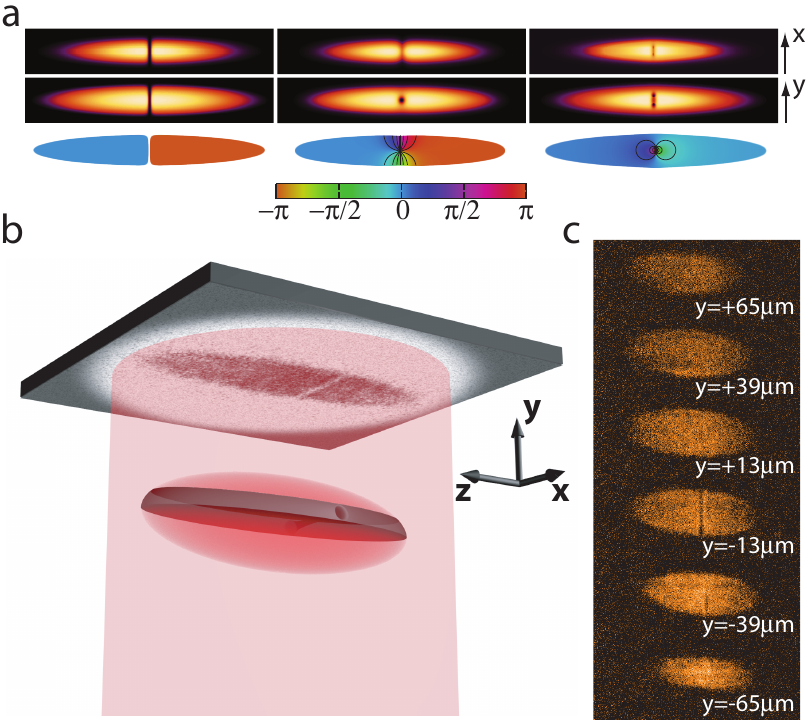}
    \caption{(a) Examples of solitary waves in 3D Bose-Einstein condensates. Shown are simulated column density profiles in the $(z-x)$ plane (upper row), the local density of the cloud in a central layer in the $(z-y)$ plane (middle row) and the phase (lower row) for a soliton (left), a solitonic vortex (center) and a vortex ring (right). The images correspond to $\mu/\hbar\omega_\perp=$7.31, 7.14, and 10.66. (b) Schematic of the experimental tomographic imaging technique. A partially masked optical pumping beam propagating along $z$ (not shown) selects a $23\,\mu$m thick slice within the expanded atom cloud for absorption imaging along the vertical $y$-direction. (c) Tomography of a unitary fermionic superfluid of $^{6}$Li atoms containing a solitary wave. Shown are density distributions of horizontal slices selected at different $y$ positions. Tomography reveals a single solitonic vortex.}
    \label{f:fig1}
    \end{center}
\end{figure}

Solitary waves are localized objects with defined energy and mass, and as such they can be described as an effective single particle emerging from a many-body environment.
Their highly localized and non-linear nature lends itself as a local probe of the medium in which they propagate.
This distinguishes them from larger-scale collective excitations such as shape oscillations of a superfluid, or from perturbative linear excitations such as phonons.
Paradigmatic examples of solitary waves in superfluids are planar solitons that separate regions of differing phase, as well as vortex rings or single vortex lines (see Fig.~\ref{f:fig1}a). The direct creation of such objects ``on demand'' in ultracold quantum gases allows for an excellent dynamical probe of novel superfluids, such as strongly interacting Fermi gases~\cite{Yefsah2013Soliton} or spin-orbit coupled Bose-Einstein condensates~\cite{Lin2011Spinorbit,Fetter2014Spinorbitvortex}.

Solitary waves vary in their degree of stability. Their energy is concentrated in the nodes of the order parameter, so the nodal plane of a soliton is energetically more costly than the nodal line of a vortex. If there is a path for decay, the system will thus tend to reduce the size of nodal regions. Planar solitons can decay via the snake instability, the undulation of the soliton plane~\cite{Muryshev1999Soliton}. For weakly interacting Bose-Einstein condensates, solitons have been created~\cite{burg99soliton,dens00} and observed to decay into vortex rings~\cite{ande01ring,dutton2001shock}. The latter further decay into a vortex-anti-vortex pair that eventually breaks up, leaving behind a single remnant solitonic vortex~\cite{Brand2002solitonicvortex, Komineas2003soliton}. The exact process was recently elucidated in a discussion of apparent soliton oscillations observed in weakly interacting BECs~\cite{Becker2008solitons,Becker2013SolitonicVortex}. Similar decay of solitons and vortex rings into single vortex lines were found recently in numerical simulations of the Ginzburg-Landau
equations~\cite{scherpelz2014vortexring}.

In a recent experiment on fermionic superfluids at MIT~\cite{Yefsah2013Soliton}, long-lived solitary waves were produced that
featured a large ratio of inertial to bare (missing) mass of over 200, evidenced by an oscillation period
over 15 times longer than the period for a single atom. The longevity as well as the large effective mass ratio were unexpected for planar solitons~\cite{Muryshev1999Soliton,Scott2011SolitonDynamics,Liao2011Solitons,Cetoli2013Snake}.
Several recent works therefore suggested that these solitary waves are vortex
rings~\cite{Bulgac2014Vortexrings,Reichl2013VortexRing,Wen2013DarkSoliton}. Vortex rings seem to be excluded for a number of reasons: Simulations required the assumption of ring sizes smaller than $20\%$ of the transverse cloud radius~\cite{Bulgac2014Vortexrings,Reichl2013VortexRing} to exhibit periods on the
order of the experimentally observed value. Such small vortex rings (see Fig.~\ref{f:fig1}a) would not lead to the straight stripes running across the entire cloud observed in the experiment. Furthermore, the formation of vortex rings would be highly sensitive to initial experimental conditions, resulting in vortex rings of varying initial size and period~\cite{Reichl2013VortexRing}. This is in contrast to the well-defined period observed over many repetitions of the experiment. However, solitons are difficult to distinguish from solitonic vortices aligned transverse to the imaging axis (see Fig.~\ref{f:fig1}a).

In this letter, we therefore investigate the nature of the long-lived solitary wave via tomographic imaging and identify the wave to be a solitonic vortex.
A (stationary) solitonic vortex obeys the same far-field phase pattern as a (stationary) dark soliton, with a phase difference of $\pi$ across the vortex. Other than resulting from soliton decay, the vortex might be directly formed in the phase imprint.
The vortex deforms the superfluid phase only in a restricted region of typical extent $R_\perp$, the transverse Thomas-Fermi cloud radius. The vortex together with the surrounding flow field thus constitutes an effective particle localized to within $R_\perp$. Its precessional motion, projected onto the long axis of the cigar-shaped
atom cloud, appears as the oscillation of a particle of inertial mass $M^*$ and bare mass $M$.
As we show below, the bare mass scales as the missing mass inside the vortex core, $M \propto m n \xi^2 R_\perp \cal{L}$, while the inertial mass $M^* \propto m n R_\perp^3/\cal{L}$ is proportional to the entire cloud volume $R_\perp^3$ in which flow is perturbed by the vortex. Here, $\xi$ is the characteristic vortex core size, $n$ is the gas density, and $\cal{L}= \ln(R_\perp/\xi)$ is a logarithmic correction that is on the order of 3 to 5 in our experiment. Thus, $M^*/M \propto R_\perp^2/\xi^2/\cal{L}$ which can easily approach 200 for our experimental parameters, thus explaining the experimental findings in~\cite{Yefsah2013Soliton}.

We create fermionic superfluids using a balanced mixture of the two lowest hyperfine states of $^6$Li, $\left|1\right\rangle$ and $\left|2\right\rangle$. A Feshbach resonance allows to tune the
interparticle interactions from the limit of Bose-Einstein condensation of tightly bound molecules towards the regime of BCS superfluidity~\cite{kett08rivista,Yefsah2013Soliton}. The atom cloud contains $1-10 \times 10^5$ atoms per spin state and is cigar-shaped due to a tight radial confinement from an optical dipole trapping beam propagating along the (horizontal) $z$-direction, in combination with a weaker, harmonic
confinement along $z$ provided by a magnetic field curvature.
The radial and axial trapping frequencies are varied in the range of $\omega_\perp/2\pi \approx 55-75\,\rm Hz$ and $\omega_z/2\pi = 5-23\,\rm Hz$. Gravity slightly distorts the trapping potential along the vertical $y$-direction, causing an anisotropy $\omega_y/\omega_x -1 \approx 5\%$.

The solitary wave is created as in~\cite{burg99soliton,dens00,Becker2008solitons,Yefsah2013Soliton}
via phase-imprinting, whereby one-half of the superfluid is exposed to a blue-detuned laser beam for a duration that causes a phase shift of the order parameter close to $\pi$. To observe the magnitude of the superfluid wavefunction,
we employ a rapid ramp to the BEC side of the Feshbach resonance during time of
flight~\cite{zwie05vort,kett08rivista,Yefsah2013Soliton}. In addition to emptying out defects such as vortex cores~\cite{zwie05vort}, the ramp effectively increases the healing length $\xi$ of the superfluid to observable values (typically $\sim 20\,\rm\mu m$). Absorption images are taken along the vertical direction (see Fig.~\ref{f:fig1}b).

In order to lift the ambiguity on the nature of the observed excitation, we employ a tomographic technique whereby only a chosen slice of the full atom cloud is imaged after time of flight (see Fig.~\ref{f:fig1}b). This method gives direct access to the local density of the 3D cloud.
Tomography is achieved by optically pumping within 10 $\mu$s all atoms outside the desired slice into hyperfine states that are off-resonant with the imaging transition for state $\left|1\right\rangle$, predominantly state $\left|6\right\rangle$.
The slice is selected by masking part of the optical pumping light with a thin wire, and projecting the wire's shadow onto the atom cloud. The slice thickness is measured to be $23(1)\,\rm\mu m$ ($=2\sigma$ of a gaussian fit), comparable to the width of the observed solitary wave after time of flight, and about one sixth of the transverse cloud diameter after expansion.

Representative tomographic images for the unitary fermionic superfluid are shown in Fig.~\ref{f:fig1}c, taken 1.6 s after the phase imprint.
A line of depletion with about 40\% contrast cuts across the entire cloud in one particular slice. This immediately demonstrates that the solitary wave is not a vortex ring. On average, only a specific one of the six slices imaged features the depletion.
The strong depletion is thus neither a planar soliton. Instead, our observation is
consistent with a single, solitonic vortex. For the present experimental conditions we observe the vortex to be horizontal in every single repetition of the experiment.
Due to the slight anisotropy of the trap, the vortex can minimize its energy by aligning along the short axis, while
orientation along the longer, intermediate axis is unstable~\cite{Svid2000VortexDynamics,Becker2013SolitonicVortex}.
Slight tilts of the vortex into the vertical direction cause partial vortex lines to be detected in a given slice, as seen for slice position $y=-39\,\rm \mu m$ in Fig.~\ref{f:fig1}c.

\begin{figure}
    \begin{center}
    \includegraphics[width=89mm]{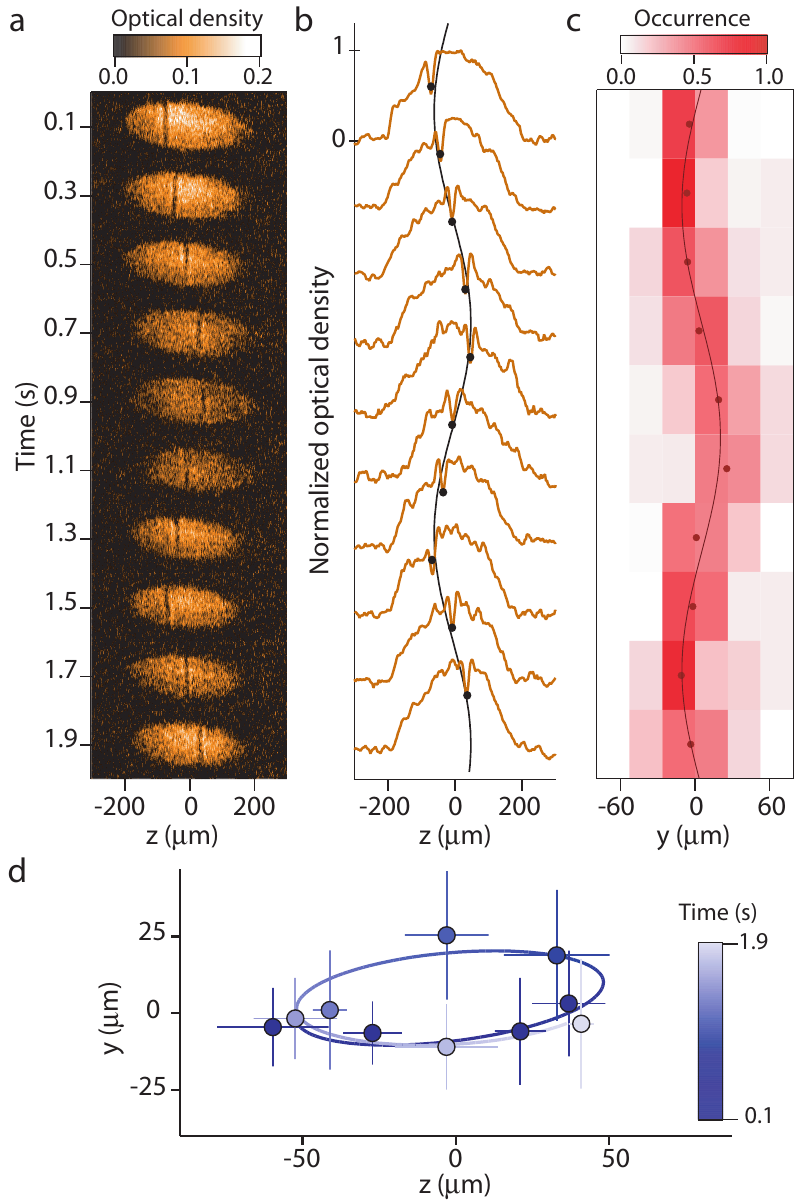}
   \caption{Observation of vortex precession in a unitary fermionic superfluid via tomographic imaging. (a) Representative horizontal slices showing the oscillation of the vortex along the $z$-axis. Time $t=0\,\rm s$ marks 600 ms after the phase imprint. The $y$ position of each slice can be inferred from c). (b) Density profiles normalized by the peak density, showing a depletion of typically 30\% contrast. The solid line is a sine fit to the vortex positions (black dots). (c) Average occurrence of the vortex at a given $y$ position of the slice as function of time, showing the vortex oscillation along the $y$-axis. Red dots: Average $y$ position of the vortex from gaussian fit at the given time; solid red line: sine fit. (d) Reconstructed precessional motion in the $z$-$y$ plane.}
    \label{f:fig2}
    \end{center}
\end{figure}

In a fully 3D setting where the radial cloud size $R_\perp$ is much larger than the vortex core size $\xi$, an
off-center transverse vortex will undergo precessional motion along equipotential
lines~\cite{Lundh2000Vortex,fetter2001vortex}. Tomographic imaging enables a measurement of the vortex position in the $z$-$y$ plane.
Representative images and column density profiles of slices containing the vortex are shown in
Fig.~\ref{f:fig2}, along with histograms of the occurrence of vortex observations in each slice.
The $z$-$y$ coordinates of the vortex lie on an ellipse with the aspect ratio of the atom cloud, as expected for
vortex precession along equipotential lines.

The period of the vortex motion can be estimated from superfluid hydrodynamics and the equation of
state in the BEC-BCS crossover~\cite{navo10eos}. Our Hamiltonian approach is analogous to that used to describe the motion of vortex rings in~\cite{Pita2013vortexring}. We take the vortex to be aligned in the transverse
$x$-direction, and located at $\vec{r}_0 = (y_0,z_0)$ in the $y$-$z$ plane.
The free energy $E_V$ of the vortex is dominated by the kinetic energy of its flow field $\vec{v} = \hbar
\nabla \phi/m_B$, where $\phi=\arctan{\frac{y-y_0}{z-z_0}}$ is the phase profile near the vortex, and $m_B =
2 m$ the boson mass. One finds $E_V \approx \int {\rm d}^3 r \frac{1}{2} m n v^2 = \frac{\pi \hbar^2
m}{m_B^2} n_{\rm 2D}(y_0,z_0) \ln(R_\perp/\xi)$ to logarithmic accuracy, i.e. in the limit
$\ln(R_\perp/\xi)\gg 1$. Here, $n$ is the gas density, $n_{\rm 2D}$ the column density along the vortex
line, $R_\perp$ is the transverse Thomas-Fermi radius, much smaller than the axial radius $R_z$, and
$\xi$ is the characteristic size of the vortex core. In the crossover we may take $\xi = \frac{1}{\sqrt{2}} \hbar/m_B c$, with $c$ the speed of sound, a definition that recovers the healing length in the BEC-regime, and yields $\xi \approx 1/ k_F$ at unitarity, a reasonable estimate~\cite{bulg03}, especially within logarithmic accuracy.

The canonical momentum of the vortex along the axial $z$-direction is given by $P_z = \int {\rm d}^3 r\,
m n v_z = \frac{m}{m_B}\int {\rm d}^3r\, \hbar n \partial_z \phi.$ Since $R_\perp \ll R_z$, the phase
gradient is concentrated in the neighborhood of the vortex in a range of size $\sim R_\perp$ along the
$z$-direction, allowing to set $n(x,y,z) \approx n(x,y,z_0)$. The integral of $\partial_z \phi$ over the
$z$-direction thus simply equals $\pi$ or $-\pi$, depending on whether the path runs along $y<y_0$ or
$y>y_0$.
One thus has $P_z  \simeq \frac{m}{m_B}\hbar \pi (\int_{-R_\perp}^{y_0} {\rm d}y -
\int_{y_0}^{R_\perp} {\rm d}y) \,n_{\rm 2D}(y,z_0) = \frac{m}{m_B}\hbar \pi \int_{-y_0}^{y_0} {\rm
d}y\,n_{\rm 2D}(y,z_0)$.
Assuming harmonic trapping and the local density approximation, we deduce the axial velocity of the
vortex from Hamilton's equation$$\dot{z}_{0} = \frac{\partial E_V}{\partial P_z} = \frac{\partial
E_V/\partial y_0}{\partial P_z/\partial y_0} =
-\frac{\omega_\perp}{\omega_z} \Omega\, y_0,$$
and similarly $\dot{y}_{0} =  \frac{\omega_z}{\omega_\perp} \,\Omega\, z_0$,
with the angular frequency
$$\frac{\Omega}{\omega_z} = \frac{2\gamma+1}{8} \frac{\hbar \omega_\perp}{\mu}
\ln\left(\frac{R_\perp}{\xi}\right).$$
Here, $\gamma \equiv \frac{\mu}{n} \frac{\partial n}{\partial \mu}$ is a polytropic
index determined by the equation of state, and $\mu$ is evaluated at the vortex position. $\gamma = 1$ in the BEC regime, while $\gamma=3/2$ at unitarity and in the BCS regime.
The equations describe the precessional motion of the vortex with angular frequency $\Omega$ along
an equipotential line of the trap with $\mu = {\rm const.}$ i.e. $y_0^2/R_\perp^2 + z_0^2/R_z^2 = {\rm
const}$.
The result is identical to what one finds by equating the Magnus force~\cite{Ao1993Magnus} $h n_{\rm
2D} \hat{x} \times \dot{\vec{r}}_0$ to the force $-\nabla E_V$ acting on the vortex, and it generalizes
the known result for vortex motion in trapped, weakly interacting Bose-Einstein
condensates~\cite{Lundh2000Vortex,fetter2001vortex} to superfluids with arbitrary equation of state.
We find the inertial mass of the vortex~\cite{Scott2011SolitonDynamics}
$$M^* = \frac{\partial P_z}{\partial \dot{z}_0} = \frac{\partial P_z/\partial y_0}{\partial
\dot{z}_0/\partial y_0}
= -\frac{4\pi}{2\gamma+1} \frac{n_{\rm 2D}R_\perp^2}{\ln(R_\perp/\xi)} m$$
which is proportional to the total mass of atoms contained in the volume $R_\perp^3$, while the bare
mass
$$M = -\frac{\partial E_V}{\partial \mu} m = - \pi\frac{2\gamma +1}{4\gamma}\, n_{\rm 2D}\xi^2
\ln\left(\frac{R_\perp}{\xi}\right) \,m$$
is only proportional to the mass of ``missing'' atoms contained in the vortex core. Here we have used
$\mu = \gamma m c^2 = \gamma \hbar^2/2m \xi^2$.
The ratio
$M^*/M \propto R_\perp^2/\xi^2 /(\ln(R_\perp/\xi))^2$ thus depends on the system size and can
become large. In contrast, the bare and inertial mass of a planar soliton are both on the order of the
mass of ``missing'' atoms in the soliton plane, $\propto n \xi R_\perp^2$, and their ratio is bound to be
on the order of unity in the crossover regime close to resonance.
Using the experimental parameters of $\mu/\hbar\omega_\perp \approx 25-35$~\cite{Yefsah2013Soliton}, the hydrodynamical model yields a normalized vortex period $T_V/T_z \approx
11-15 $  and effective mass ratio $M^*/M =130-220$, in close agreement with the measured values.

\begin{figure}
    \begin{center}
    \includegraphics[width=89mm]{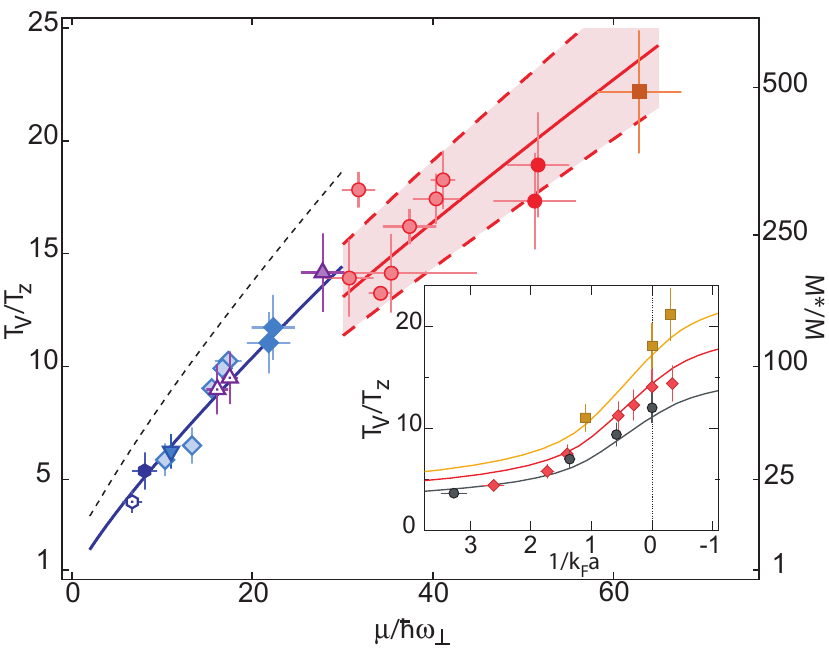}
  \caption{Normalized period of the solitonic vortex $T_V/T_z$, as a function of the normalized chemical potential $\mu/\hbar\omega_{\perp}$. Experimental data are for magnetic fields $B=850\,\rm G$ (BCS side, square), $832\,\rm G$ (unitarity, circle), and for the BEC-side at $800\,\rm G$ (triangle), $760\,\rm G$ (diamond), $740\,\rm G$ (inverted triangle), and $700\,\rm G$ (hexagon). $\omega_z/2\pi$ was $23\,\rm Hz$ (solid symbols), $10\,\rm Hz$ (framed), $5\,\rm Hz$ (dotted). Predictions in the BEC regime: Solid blue line from~\cite{Lundh2000Vortex}, dashed line from~\cite{fetter2001vortex}. Red solid curve: hydrodynamic prediction on resonance, assuming $\xi=1/k_F$. The error band is bounded by the choice $\xi=1/2k_F$ (lower bound) and $\xi=2/k_F$ (upper bound). Inset: $T_V/T_z$ as a function of the interaction parameter $1/k_Fa$, data from \cite{Yefsah2013Soliton}. Square, diamond, and circle are for $\omega_z/2\pi= 23$, $10$, and $5\,\rm Hz$, respectively. Atom numbers range from $N/2 = 1 \times 10^5$ per spin state in the BEC regime to $3\times 10^5$ around resonance. Solid curves: hydrodynamic prediction fixing $N/2=3\times 10^5$, and $\omega_z/2\pi= 23\,\rm Hz$ (gold), $10\, \rm Hz$ (red), and $5\,\rm Hz$ (black).}
    \label{f:fig3}
    \end{center}
\end{figure}

We have taken extensive data for the vortex period in the BEC-BCS
crossover exploring a wide range of chemical potentials. Fig.~\ref{f:fig3} shows the normalized period
$T_V/T_z$ versus $\mu/\hbar\omega_\perp$ including data for several aspect ratios, interaction
strengths and atom numbers. Chemical potentials were extracted from the measured axial
Thomas-Fermi radius of the cloud and the known axial trapping frequency.
The inset in Fig.~\ref{f:fig3} shows the data from~\cite{Yefsah2013Soliton} along with the theoretical prediction for a fixed, characteristic atom number of $N/2 = 3\times 10^5$ per spin state, using the known equation of state in the BEC-BCS crossover~\cite{navo10eos}.
The data is in good agreement with the approximate theory, from the BEC regime towards
resonance and into the BCS regime.
Corrections beyond logarithmic accuracy could be important as $\ln(R_\perp/\xi)$ is only $3-5$, but they are not known in the crossover beyond the weakly interacting BEC regime, and are subject of debate~\cite{Sonin2013VortexMass}. Generally, there will be a contribution to the vortex' inertial mass from superfluid backflow, the Baym-Chandler mass~\cite{Baym1982VortexMass}. For a strongly interacting Bose gas, quantum depletion localized inside vortex cores will modify the inertial and bare mass~\cite{Fetter1971vortexcore,Fetter1972QuantumDepletion}. In the BCS regime, one expects a contribution due to fermions trapped in Andreev bound states inside the vortex core~\cite{caro64bound}, the Kopnin mass~\cite{Kopnin1978Mass}.
For the Gross-Pitaevskii equation describing weakly interacting BECs, the vortex period was found in
numerical calculations to be well-described by the approximate formula even when the transverse cloud
size became comparable to the size of a vortex~\cite{parker2004thesis}.
For a molecular BEC, the prediction from the GP equation is~\cite{Lundh2000Vortex,fetter2001vortex} $\frac{\Omega}{\omega_z} = \frac{3}{8} \frac{\hbar \omega_\perp}{\mu} \left(\ln\left(\frac{R_\perp}{\xi}\right) + \frac{3}{4}\right)$ and is shown in Fig.~\ref{f:fig3} to agree well with the data.

An interesting future investigation concerns the early times a few milliseconds after the phase
imprinting. Is the single observed vortex a result of multiple decay processes, in which an initial planar
soliton decays into a vortex ring, that further decays into vortex-anti-vortex pairs, followed by a
``sling-shot'' event~\cite{Becker2013SolitonicVortex} by which one of the vortices is ejected? Or does
the phase imprint rather directly create vortices of a given circulation? For example, solitons that are slightly tilted with respect to the transverse direction can efficiently
convert into solitonic vortices of one type of charge, removing the required angular momentum from a collective mode of the gas cloud~\cite{parker2004thesis}.

In conclusion, we have implemented a tomographic imaging technique that allowed to conclusively
demonstrate that a long-lived solitary wave observed in our fermionic superfluid is a solitonic vortex.
The vortex is topologically protected, explaining the long lifetime of the wave, and its theoretical inertial
to bare mass ratio agrees with that found experimentally.
Solitonic vortices can be expected to occur as
persistent defects created via a Kibble-Zurek
mechanism~\cite{weil08,Lamporesi2013KibbleZurekSoliton}, via
phase-imprinting~\cite{Becker2008solitons,Becker2013SolitonicVortex} or even via thermal excitations,
as hinted at by the observation of thermally induced defects in~\cite{Yefsah2013Soliton}.
They also correspond to the ``N''-shaped vortices created via rotation in~\cite{Rosenbusch2002vortex}, in the limit of zero rotation frequency (called ``S''-shaped in~\cite{Komineas2005Svortex}).
Further studies on this topological excitation created ``on demand'' concern the interaction of multiple solitonic
vortices in fermionic superfluids, a measurement of the current-phase relation of solitonic
vortices~\cite{Komineas2003soliton}, their contribution to flow resistance of the
superfluid~\cite{Halperin2010Resistance} and the observation of Andreev states bound to vortex
cores~\cite{caro64bound}.

We would like to thank David Huse and Lev Pitaevskii for fruitful discussions. This work was supported
by the NSF, the ARO MURI on Atomtronics, AFOSR PECASE, ONR, a grant from the Army Research Office
with funding from the DARPA OLE program and the David and Lucile Packard Foundation.


%Merlin.mbs v4.21 2009-07-09.
%

\end{document}